\documentclass[aps,amsfonts,nofootinbib,preprintnumbers,nobalancelastpage]{revtex4}
\usepackage{graphicx}
\usepackage{epsfig}
\usepackage{xcolor}
\usepackage{rotating}
\usepackage{amsmath}
\usepackage{braket}

\newcommand{\cw}{{\rm c_W} }
\newcommand{\sw}{{\rm s_W} }

\newcommand{\cwsq}{{\rm c_W^2} }
\newcommand{\swsq}{{\rm s_W^2} }

\newcommand{\swt}{{\rm s_W^3}}

\newcommand{\swf}{{\rm s_W^4}}

\def\ctw{c_{2W}}
\def\stw{s_{2W}}

\newcommand{\na}{{\rm \--}}
\newcommand{\fw}{{f_{\rm W}}}
\newcommand{\fb}{{f_{\rm B}}}
\newcommand{\nc}{{{\rm N}_{\rm c}}}

\begin{document}
\hfill{YITP-SB-14-46}

\title{Unitarity Constraints on Dimension-Six Operators}
\author{Tyler Corbett}
\email{corbett.t.s@gmail.com}
\affiliation{%
  C.N.~Yang Institute for Theoretical Physics, SUNY at Stony Brook,
  Stony Brook, NY 11794-3840, USA}

\author{O.\ J.\ P.\ \'Eboli}
\email{eboli@fma.if.usp.br}
\affiliation{Instituto de F\'{\i}sica,
             Universidade de S\~ao Paulo, S\~ao Paulo -- SP, Brazil.}
\author{M.\ C.\ Gonzalez--Garcia} \email{concha@insti.physics.sunysb.edu}
\affiliation{%
  Instituci\'o Catalana de Recerca i Estudis Avan\c{c}ats (ICREA),}
\affiliation {Departament d'Estructura i Constituents de la Mat\`eria, 
Universitat
  de Barcelona, 647 Diagonal, E-08028 Barcelona, Spain}
\affiliation{%
  C.N.~Yang Institute for Theoretical Physics, SUNY at Stony Brook,
  Stony Brook, NY 11794-3840, USA}
\begin{abstract}
 We obtain the partial-wave unitarity constraints on dimension-six
 operators stemming from the analyses of vector boson and Higgs
 scattering processes as well as the inelastic scattering of
 standard model fermions into electroweak gauge bosons. We take into
 account all coupled channels, all possible helicity amplitudes, and
 explore a six-dimensional parameter space of anomalous couplings.
Our analysis shows that for those operators
   affecting the Higgs couplings, present 90\% confidence level constraints from global
   data analysis of Higgs and electroweak data are such that unitarity
   is not violated if $\sqrt{s}\leq 3.2\;{\rm TeV}$.  For the purely
   gauge-boson operator $O_{WWW}$, the present bounds from
   triple-gauge boson analysis indicate that within its presently
   allowed 90\% confidence level range unitarity can be violated in $f\bar f' \to V V'$
   at center-of-mass energy $\sqrt{s}\geq 2.4\;{\rm TeV}$.  
\end{abstract}
\maketitle
\renewcommand{\baselinestretch}{1.15}
\section{Introduction}
The standard model (SM) of electroweak interactions has been extremely
successful in the description of the available data, and up to now
there is no clear experimental evidence that challenges its
predictions.  As long as no new state has been observed, effective
lagrangians provide a well defined systematic way to parametrize
departures from the standard model.  Furthermore, the recent discovery
of a particle resembling a light Higgs boson indicates that the
$SU(2)_L \otimes U(1)_Y$ gauge symmetry might be realized linearly in
the effective theory. Therefore, we can parametrize the effects of new
physics by adding to the SM lagrangian higher dimension operators made
up of the SM fields.  Within the global symmetries of the SM the
lowest dimension of the new operators is six, hence we include those
dimension-six operators:
\begin{equation}
\mathcal{L}_{\rm eff}= \mathcal{L}_{\rm SM} + \sum_{n}
\frac{f_n}{\Lambda^2}\mathcal{O}_n
\label{eq:leff}
\end{equation}
where, in general, the dimension-six operators $\mathcal{O}_n$ involve
gauge-bosons, Higgs doublets, fermionic fields, and (covariant-)
derivatives of these fields. Each operator has a corresponding
coupling $f_n$ and $\Lambda$ is the characteristic energy scale at
which new physics (NP) becomes apparent. 

It is well known that nonrenormalizable higher dimensional operators
give rise to rapid growth of the scattering amplitudes with energy,
leading to partial-wave unitarity violation. This fact constrains the
energy range where the low energy effective theory is valid once the
coefficients $f_n$ are fixed. With this aim in mind in this work we
revisit the bounds from partial-wave unitarity on $\mathcal{L}_{\rm
  eff}$ arising from vector boson and Higgs boson scattering, as well
as inelastic processes $f\bar f' \to V V'$ where $f^{(\prime)}$ is a SM fermion
and $V^{(\prime)}$ is an electroweak gauge boson.

Previous works in the literature studied the unitarity bounds on some
of the dimension-six operators either considering only one
non-vanishing coupling at a time, and/or they did not take into
account coupled channels, or they worked in the framework of effective
vertices ~\cite{Bilchak:1987cp, Gounaris:1993fh, Gounaris:1994cm,
  Gounaris:1995ed, Degrande:2013mh,Baur:1987mt}. Here, we complete
these previous analyses by considering the effects of coupled channels
leading to the strongest constraints, including both elastic and
inelastic channels, and also by analyzing the general six-dimensional
parameter space of relevant anomalous couplings. Moreover, we only
consider contributions up to order $1/\Lambda^2$ to apply
systematically the effective field theory approach.

The outline of this article is as follows. We summarize the formalism
employed in Sec.\ref{sec:forma}, in particular Sec.\ref{sec:lag}
contains the basis of operators considered, and in Sec.\ref{sec:unit} we
briefly present the standard partial-wave unitarity constraints from
elastic gauge boson scattering and inelastic $f\bar f' \to V V'$ processes.
Section \ref{sec:results} contains our results from the unitarity
analysis and we compare those with the presently allowed range from
collider searches.  In particular we conclude that, even in the most
general case, those operators affecting the Higgs couplings do not
violate unitarity for center-of-mass energies $\sqrt{s}\leq 3.2\;{\rm
  TeV}$ within the range presently allowed from global data analysis
of Higgs and electroweak data at 90\% CL.

\section{Analyses framework}
\label{sec:forma}
In this section we present the effective interactions considered in
this work, as well as the unitarity relations that we use to constrain
them

\subsection{Effective Lagrangian}
\label{sec:lag}

We parametrize deviations from the Standard Model (SM) in terms of
dimension-six effective operators as in Eq.~(\ref{eq:leff}).  The
dimension-six basis contains 59 independent operators, up to flavor
and Hermitian conjugation, which are sufficient to generate the most
general S-matrix elements given the SM gauge symmetry and that baryon
and lepton number symmetries are obeyed by the NP~\cite{59ops}.
Exploiting the freedom in the choice of basis, we work in that of
Hagiwara, Ishihara, Szalapski, and Zeppenfeld (HISZ)
\cite{HISZ,HISZ2}. 

In what follows we consider bosonic operators relevant to two-to-two
scattering processes involving Higgs and/or gauge bosons at tree
level, and will impose $C$- and $P$-evenness on the operators, which
leaves us with ten dimension-six operators. These operators can be
classified into three groups according to their field
content\footnote{We do not consider operators with higher derivative
  kinetic term for the Higgs and gauge bosons.  They can be traded by
  a combination of the operators considered plus some fermionic
  operators and hence do not lead to new unitarity violating effects
  in the scattering amplitudes studied here; see \cite{Brivio:2014pfa}
  for the explicit derivation for the case of the operator with a higher
  derivative kinetic term for the Higgs.}
\begin{itemize}

\item pure gauge operators, in this class there is just one operator
\begin{equation}
  \mathcal{O}_{WWW} = {\rm Tr}[\widehat{W}_{\mu}^{\nu}
    \widehat{W}_{\nu}^{\rho}\widehat{W}_{\rho}^{\mu}] \;\;;
\label{eq:www}
\end{equation}
\item gauge-Higgs operators which include
\begin{eqnarray}
  \mathcal{O}_{WW}		
&=& \Phi^\dagger\widehat{W}_{\mu\nu}\widehat{W}^{\mu\nu}\Phi \;\;, \\
  \mathcal{O}_{BB}		
&=& \Phi^\dagger\widehat{B}_{\mu\nu}\widehat{B}^{\mu\nu}\Phi  \;\;,	\\
  \mathcal{O}_{BW}		
&=& \Phi^\dagger\widehat{B}_{\mu\nu}\widehat{W}^{\mu\nu}\Phi  \;\;,	\\
  \mathcal{O}_{W}		
&=&	(D_\mu\Phi)^\dagger\widehat{W}^{\mu\nu}(D_\nu\Phi)  \;\;,	\\
  \mathcal{O}_{B}		
&=&	(D_\mu\Phi)^\dagger\widehat{B}^{\mu\nu}(D_\nu\Phi)  \;\;,	\\
  \mathcal{O}_{\Phi,1}		
&=&	(D_\mu\Phi)^\dagger\Phi\Phi^\dagger(D^\mu\Phi) \;\;,	\\
  \mathcal{O}_{\Phi,4}		
&=&	(D_\mu\Phi)^\dagger(D^\mu\Phi)(\Phi^\dagger\Phi)  \;\;;
\end{eqnarray}
\item and pure Higgs operators:
\begin{eqnarray}
  \mathcal{O}_{\Phi,2} &=&
  \frac{1}{2}\partial^\mu(\Phi^\dagger\Phi)\partial_\mu(\Phi^\dagger\Phi)  \;\;,
\label{eq:phi2} \\
  \mathcal{O}_{\Phi,3} &=&\frac{1}{3}(\Phi^\dagger\Phi)^3  \;\;,
\label{eq:phi3}
\end{eqnarray}
\end{itemize}
where $\Phi$ stands for the Higgs doublet and we have adopted the
notation $\widehat{B}_{\mu\nu}\equiv i(g^\prime/2)B_{\mu\nu}$,
$\widehat{W}_{\mu\nu}\equiv i(g/2)\sigma^aW^a_{\mu\nu}$, $g$ with
$g^\prime$ being the $SU(2)_L$ and $U(1)_Y$ gauge couplings
respectively, and $\sigma^a$ the Pauli matrices.

The dimension-six operators given in
Eqs.~(\ref{eq:www})--(\ref{eq:phi3}) modify the triple and quartic
gauge boson couplings, the Higgs couplings to fermions and gauge
bosons, and the Higgs self-couplings.; see
Table~\ref{tab:coupl}. Further details are presented in the
appendix~\ref{app:ano}. A thorough discussion of the effects of the
operators relevant to Higgs physics and anomalous gauge couplings in
the basis here employed can be found in \cite{us1, us2, us3}.

\begin{table}
\begin{tabular}{|c|c|c|c|c|c|c|c|c|}
\hline 
& $VVV$ & $VVVV$ & $HVV$ & $HVVV$ & $HHVV$ &
$HHH$ & $HHHH$ & $H\bar{f}f$ 
\\ \hline 
$\mathcal{O}_{WWW}$ & X & X & & & & & & 
\\ \hline 
$\mathcal{O}_{WW}$ & & & X & X & X & & & 
\\ \hline
$\mathcal{O}_{BB}$ & & & X & & X & & & 
\\ \hline 
$\mathcal{O}_{BW}$ & X & X & X & X & X & & & 
\\ \hline 
$\mathcal{O}_{W}$ & X & X & X & X & X & & & 
\\ \hline 
$\mathcal{O}_{B}$ & X & & X & X & X & & & 
\\ \hline
$\mathcal{O}_{\Phi,1}$ & X  &X & X & & X & X & X & X 
\\ \hline
$\mathcal{O}_{\Phi,2}$ & & & X & & X & X & X & X 
\\ \hline
$\mathcal{O}_{\Phi,3}$ & & & & & & X & X & 
\\ \hline
$\mathcal{O}_{\Phi,4}$ & & & X & & X & X & X & X 
\\\hline
\end{tabular}
\caption{Couplings relevant for our analysis that are modified by the
  dimension--six operators in
  Eqs.~(\ref{eq:www})--(\ref{eq:phi3}). Here, $V$ stands for any
  electroweak gauge boson, $H$ for the Higgs and $f$ for SM fermions.}
\label{tab:coupl}
\end{table}

We notice first that operators $\mathcal{O}_{BW}$ and
$\mathcal{O}_{\phi,1}$ contribute at tree level to the oblique
electroweak precision parameter $T$ (or
$\Delta\rho$)~\cite{Hagiwara:1986vm, DeRujula:1991se, Hagiwara:1993ck,
  Alam:1997nk} . Taking into account that the present available data
impose strong bounds on these parameters~\cite{Baak:2014ora}, the
couplings $f_{BW}$ and $f_{\Phi,1}$ are severely constrained,
consequently we neglect $\mathcal{O}_{BW}$ and $\mathcal{O}_{\phi,1}$
in our analyses.  This leaves us with a basis of 8
operators. Furthermore for large center--of--mass energies
($\sqrt{s}$), which we will take to mean $ \sqrt{s} \gg M_{W,Z,H}$ for
our analysis, the behavior of $\mathcal{O}_{\phi,2}$ and
$\mathcal{O}_{\phi,4}$ is the same up to a sign for the scattering
processes considered and as such for our discussion we can quantify
their behavior by a single operator coefficient:
\begin{equation}\label{phi24relation}
\frac{f_{\Phi2,4}}{\Lambda^2}\equiv \frac{f_{\Phi,2}-f_{\Phi,4}}{\Lambda^2}
\;\; .
\end{equation} 
This is expected since $\mathcal{O} _{\Phi, 2}+\mathcal{O} _{\Phi, 4}$
can be traded via equations of motion by a combination of Yukawa-like
operators which do not contribute to the $2\rightarrow 2$ scattering
processes considered.

Additionally $\mathcal{O}_{\Phi,3}$ modifies the Higgs
self-couplings as well as the relation between the Higgs mass, its vev
and the potential term $\lambda$ (see Appendix A).  However these
effects do not induce unitarity violation in the $2\rightarrow 2$
scattering processes.

In summary our study will be carried out in terms of the six relevant
operator coefficients $f_{W}$, $f_{B}$, $f_{WW}$, $f_{BB}$, $f_{WWW}$,
and $f_{\Phi 2,4}$.

\subsection{Partial-wave unitarity}
\label{sec:unit}

In the two-to-two scattering of electroweak gauge bosons ($V$)
\begin{equation}
{V_1}_{\lambda_1}{V_2 }_{\lambda_2} \to {V_3}_{\lambda_3}{V_4}_{\lambda_4}
\end{equation}
the corresponding helicity amplitude can be expanded in partial waves
in the in the center--of--mass system as \cite{Jacob:1959at}
\begin{equation}
\mathcal{M}
({V_1}_{\lambda_1}{V_2 }_{\lambda_2} \to {V_3}_{\lambda_3}{V_4}_{\lambda_4}) 
=16 \pi \sum_J
\left ( J+\frac{1}{2} \right)~ 
\sqrt{1+\delta_{{V_1}_{\lambda_1}}^{{V_2}_{\lambda_2}}}
\sqrt{1+\delta_{{V_3}_{\lambda_3}}^{{V_4}_{\lambda_4}}}
d_{\lambda\mu}^{J}(\theta) ~e^{i M \varphi}
~ T^J({V_1}_{\lambda_1}{V_2 }_{\lambda_2} \to {V_3}_{\lambda_3}{V_4}_{\lambda_4}) 
\;\;,
\label{eq:helamp}
\end{equation}
where $\lambda=\lambda_1-\lambda_2$, $\mu=\lambda_3-\lambda_4$, $M =
\lambda_1 - \lambda_2 - \lambda_3 + \lambda_4$, and $\theta$
($\varphi$) is the polar (azimuth) scattering angle. $d$ is the usual
Wigner rotation matrix. In the case one of the vector bosons is
replaced by the Higgs we can still employ this expression by setting
the correspondent $\lambda$ to zero.

Partial-wave unitarity for the elastic channels requires that 
\begin{equation}
|T^J({V_1}_{\lambda_1}{V_2 }_{\lambda_2} \to {V_1}_{\lambda_1}{V_2}_{\lambda_2}) 
| \le 2 \;\;,
\label{eq:unitcond}
\end{equation}
where we have assumed $s\gg (M_{V_1}+M_{V_2})^2$.  More stringent
bounds can be obtained by diagonalizing $T^J$ in the particle and
helicity space and then applying the condition in
Eq.~(\ref{eq:unitcond}) to each of the eigenvalues.

We have also studied unitarity constraints from fermion annihilation
processes \cite{Baur:1987mt}
\begin{equation}
{f_1}_{\sigma_1}{\bar {f_2}}_{\sigma_2} \to 
{V_3}_{\lambda_3}{V_4}_{\lambda_4} \; .
\label{eq:ffVV}
\end{equation}
In this case the partial-wave expansion is given by
\begin{equation}
\mathcal{M}
({f_1}_{\sigma_1}{\bar {f_2}}_{\sigma_2} \to 
{V_3}_{\lambda_3}{V_4}_{\lambda_4})
 =16 \pi \sum_J
\left ( J+\frac{1}{2} \right)~\delta_{\sigma_1, -\sigma_2}
d_{\sigma_1-\sigma_2,\lambda_3-\lambda_4}^{J}(\theta)
~ T^J({f_1}_{\sigma_1}{\bar {f_2}}_{\sigma_2} \to 
{V_3}_{\lambda_3}{V_4}_{\lambda_4}) \;\;,
\label{eq:helamp2}
\end{equation}
where, for simplicity, we have set $\varphi=0$.  These processes
proceed via s-channel exchange of a $J=1$ vector boson and therefore
in the limit of massless fermions those must appear in opposite
helicity states, a condition which is explicitly enforced in the
expression above by the inclusion of the term $\delta_{\sigma_1,
  -\sigma_2}$.

Following the procedure presented in Ref.~\cite{Baur:1987mt} the
unitarity bound on the inelastic production of gauge boson pairs in
Eq.~(\ref{eq:ffVV}) is found by relating the corresponding amplitude to that of
the elastic process
\begin{equation}
{f_1}_{\sigma_1}{\bar {f_2}}_{\sigma_2} \to 
{f_1}_{\sigma_1}{\bar {f_2}}_{\sigma_2}   \;.
\end{equation}
In this case the unitarity relation is given by
\begin{eqnarray}
2{\rm Im}[T^J({f_1}_{\sigma_1}{\bar {f_2}}_{\sigma_2} 
\to {f_1}_{\sigma_1}{\bar {f_2}}_{\sigma_2})]&=&
\left|T^J({f_1}_{\sigma_1}{\bar {f_2}}_{\sigma_2}\to {f_1}_{\sigma_1}{\bar {f_2}}_{\sigma_2})\right|^2 \label{eq:unitff}\\ 
&&+\sum_{{V_3}_{\lambda_3},{V_4}_{\lambda_4}}
\left|T^J({f_1}_{\sigma_1}{\bar {f_2}}_{\sigma_2}\to 
{V_3}_{\lambda_3}{V_4}_{\lambda_4})\right|^2+
\sum_N
\left|T^J({f_1}_{\sigma_1}{\bar {f_2}}_{\sigma_2}\to N)\right|^2 \: ,\nonumber
\end{eqnarray}
where as before we take the limit $s\gg (M_{V_1}+M_{V_2})^2$. $N$
represents any state into which ${f_1}_{\sigma_1}{\bar {f_2}}_{\sigma_2}$
can annihilate which also does not consists of two gauge
bosons. Eq.~(\ref{eq:unitff}) is a quadratic equation for ${\rm
  Im}[T^J({f_1}_{\sigma_1}{\bar {f_2}}_{\sigma_2} \to
  {f_1}_{\sigma_1}{\bar {f_2}}_{\sigma_2})]$ which only admits a
solution if
\begin{equation}
\sum_{{V_3}_{\lambda_3},{V_4}_{\lambda_4}}
\left|T^J({f_1}_{\sigma_1}{\bar {f_2}}_{\sigma_2}\to 
{V_3}_{\lambda_3}{V_4}_{\lambda_4})\right|^2 \leq 1 \; . 
\label{eq:unitcond2}
\end{equation}
The strongest bound can be found by considering some optimized linear
combination
\begin{equation}
   | X \rangle = \sum_{f_1,\sigma_1} x_{f_2,\sigma_2} 
     | {f_1}_{\sigma_1} {\bar{f_2}}_{\sigma_2} \rangle
\end{equation}
with the normalization condition $\sum_{f\sigma} | x_{f\sigma}|^2=1$,
for which the amplitude $T^J(X\to {V_3}_{\lambda_3}{V_4}_{\lambda_4})$
is largest.

\section{Results}
\label{sec:results}

Let us start by considering all two-to-two Higgs and electroweak
gauge-boson scattering processes.  We have calculated the scattering
amplitudes for all possible combinations of particles and helicities
generated by the SM extended with the dimension-six operators
presented in Sec.\ref{sec:lag}.  In doing so we have consistently kept
the anomalous terms induced by the dimension-six terms in linear
order.  It is interesting to notice that to this order there is no
amplitude that diverges as $s^2$.  This is a result of gauge invariance
enforcing that the corresponding triple and quartic vertices satisfy
the requirements for the cancellation of the $s^2$ terms to take
place~\cite{Csaki:2003dt}.

All together we find 26 processes (in particle space) which yield some
helicity amplitude that grows as $s$ for some of the dimension-six
operators while the rest are constant or vanishing at large
energies. We give the corresponding expressions of the parts of the 
amplitudes which grow as $s$  in Tables \ref{tab:fphis}--\ref{tab:fwww}.
Table~\ref{tab:fphis} displays the terms in the amplitudes that grow
as $s$ at high energies due to the contributions of the operators
$\mathcal{O}_{\Phi,4}$ and $\mathcal{O}_{\Phi,2}$. It is interesting
to notice that these operators lead to unitarity violation only for
the scattering of longitudinal gauge bosons. This is expected as
these operators do not generate higher derivative terms beyond those
already present in the SM in the triple and quartic couplings.  The
amplitudes that violate unitarity at high energies due to the presence
of $\mathcal{O}_{W}$ ($\mathcal{O}_{B}$) are presented in
Table~\ref{tab:fw} (\ref{tab:fb}), the results for $\mathcal{O}_{WW}$
and $\mathcal{O}_{BB}$ are contained in Table~\ref{tab:fwwfbb}, and 
those for $\mathcal{O}_{WWW}$ are shown in Table~\ref{tab:fwww}.  As
we can see from these tables, for these five operators the growth as
$s$ of the amplitudes occurs not only for the scattering of
longitudinal gauge bosons but also for transversely polarized ones.
Notice also that all amplitudes which grow with $s$ generated by
$\mathcal{O}_{\Phi,4}$, $\mathcal{O}_{\Phi,2}$, $\mathcal{O}_{W}$,
$\mathcal{O}_{B}$, $\mathcal{O}_{WW}$, and $\mathcal{O}_{BB}$ have
only $J=0$ or $J=1$ partial-wave projections.  $\mathcal{O}_{WWW}$
leads to violation of unitarity also in helicity amplitudes with
projections over $J\geq 2$ partial waves. Notwithstanding, as the
bounds are weakened for increasing $J$, we compute the constraints
using only the amplitudes in $J=0$ and $J=1$ partial waves.

With the results in Tables \ref{tab:fphis}--\ref{tab:fwww} we proceed
to build the $T^0$ and $T^1$ amplitude matrices in particle and
parameter space. These matrices are formed with the s-divergent
amplitudes corresponding to all combinations of gauge boson and Higgs
pairs with a given total charge $Q=2,1,0$ with possible projections on
a given partial wave $J$ which are:
\begin{equation}
\begin{array}{c | cccccccccc r}
(Q,J)			&{\rm States}	&&&&&&&&&& \rm{Total} \\
\hline
\hline
(2,0) 		&	W^+_\pm W^+_\pm		&W^+_0W^+_0
&&&&&&&&&3\\[+0.1cm]
\hline
(2,1) 		& 	W^+_\pm W^+_\pm		&W^+_\pm W^+_0 	&W^+_0W^+_\pm
&&&&&&&&6	\\[+0.1cm]
\hline
(1,0) 		&	W^+_\pm Z_\pm		&W^+_0Z_0		
		&W^+_\pm \gamma_\pm		&W^+_0H		
&&&&&&&6\\[+0.1cm]
\hline
(1,1) 		&	W^+_0Z_0		&W^+_\pm Z_0	&W^+_0Z_\pm
	&W^+_\pm Z_\pm	&	W^+_0\gamma_\pm	&W^+_\pm\gamma_\pm
		&W^+_0H					&W^+_\pm H	
&&&14\\ [+0.1cm]
\hline
(0,0) 		&	W^+_\pm W^-_\pm		&W^+_0W^-_0		
		&Z_\pm Z_\pm	&Z_0Z_0	  	&	Z_\pm\gamma_\pm		
&\gamma_\pm\gamma_\pm	&Z_0H	&HH		
&&&12\\[+0.1cm]
\hline
(0,1)		&	W^+_0W^-_0			&W^+_\pm W^-_0
		&W^+_0W^-_\pm	&W^+_\pm W^-_\pm	&	Z_\pm Z_0
		&Z_0Z_\pm 				&Z_0\gamma_\pm	
   &	Z_0H				&Z_\pm H	&\gamma_\pm H	
&18\\[+0.1cm]
\end{array}
\label{eq:blocks}
\end{equation}
where upper indices indicate charge and lower indices helicity,  
and taking into account the relation 
\begin{equation}
 T^J({V_1}_{\lambda_1}{V_2 }_{\lambda_2} \to {V_3}_{\lambda_3}{V_4}_{\lambda_4})
=(-1)^{\lambda_1-\lambda_2-\lambda_3+\lambda_4}
  T^J({V_1}_{-\lambda_1}{V_2 }_{-\lambda_2} \to {V_3}_{-\lambda_3}{V_4}_{-\lambda_4}) 
\;\;.
\end{equation}
We present in the right-hand side of Eq.~(\ref{eq:blocks}) the
dimensionality of the corresponding $T^J$. For example for $Q=2$,
$T^{0}$ in the basis
$\left(W^+_{+}W^+_{+},W^+_{0}W^+_{0},W^+_{-}W^+_{-}\right)$ is the
$3\times 3$ matrix\footnote{Notice that we have introduce in
  Eq.~(\ref{eq:helamp}) the symmetry factors
  $\sqrt{1+\delta_{{V_1}_{\lambda_1}}^{{V_2}_{\lambda_2}}}$ and
  $\sqrt{1+\delta_{{V_3}_{\lambda_3}}^{{V_4}_{\lambda_4}}}$ in the
  definition of the corresponding $T^J$ amplitude while in some other
  conventions they are included in the definition of the two equal
  gauge boson states.}
\begin{equation}
\frac{1}{8\pi} s\left(
\begin{array}{ccc}
0 &	0&	\frac{3}{\swsq} e^4 f_{WWW}  \\
0	& 
-\frac{3}{8\cwsq} e^2 f_B-\frac{3}{8\swsq} e^2 f_W-
\frac{1}{2} f_{\Phi_{2,4}} & 0 \\
\frac{3}{\swsq} e^4 f_{WWW}  &0&0 \\
\end{array}
\right)  \;\;.
\end{equation}   

In order to obtain the most stringent bounds on the coefficients
$f_n/\Lambda^2$ we diagonalize the six $T^J$ matrices and impose the
constraint Eq.~(\ref{eq:unitcond}) on each of their eigenvalues. We
find that there are 50 possible nonzero eigenvalues of the total 59.
Considering only one operator different from zero at a time, we find
that the strongest constraint arise from the following eigenvalues:
\begin{eqnarray}
\left|\frac{3}{16\pi}\frac{f_{\Phi2,4}}{{\Lambda^2}}s\right|	
&\le	&2 
\;\;\;  \Rightarrow \;\;\; 
\left |\frac{f_{\Phi2,4}}{{\Lambda^2}}s\right| \le 33	\;\;,
\nonumber\\ 
\left|1.4\frac{g^2}{8\pi}\frac{f_{W}}{\Lambda^2}s\right|	&\le&2
\;\;\;  \Rightarrow \;\;\; 
\left|\frac{f_{W}}{\Lambda^2}s\right| \le 87 \;\;,
\nonumber \\
\left|\frac{g^2\sw(\sqrt{9+7\cwsq}+3\sw)}{128\cwsq\pi}\frac{f_B}{\Lambda^2}s\right|		&\le	&2
\;\;\;  \Rightarrow \;\;\; 
\left|\frac{f_{B}}{\Lambda^2}s\right| \le 617
\nonumber \\
\left|\sqrt{\frac{3}{2}}\frac{g^2}{8\pi}\frac{f_{WW}}{\Lambda^2}s\right| &
\le	&2
\;\;\;  \Rightarrow \;\;\; 
\left|\frac{f_{WW}}{\Lambda^2}s\right| \le 99 \;\;, \label{eq:vvbounds}\\
\left|.20\frac{g^2}{8\pi}\frac{f_{BB}}{\Lambda^2}s\right|	&\le&2
\;\;\;  \Rightarrow \;\;\; 
\left|\frac{f_{BB}}{\Lambda^2}s\right| \le 603 \;\;,
 \nonumber \\
\left|(1+\sqrt{17-16\cwsq\swsq})\frac{3g^4}{32\pi}\frac{f_{WWW}}{\Lambda^2}s\right|&\le&2 
\;\;\;  \Rightarrow \;\;\; 
\left|\frac{f_{WWW}}{\Lambda^2}s\right| \le 85 \;\;.
\nonumber 	
\end{eqnarray}

Next we consider the effects of $\mathcal{O}_W$, $\mathcal{O}_B$, and
$\mathcal{O}_{WWW}$ on fermion scattering into gauge bosons pairs.
They are due to the induced  modification of the triple gauge boson 
couplings. We considered the  total charge $Q=0$ processes
\[
l\bar{l}\to W^+W^- \;\;\;,\;\;\; 
\nu\bar{\nu}\to W^+W^- \;\;\;\hbox{and}\;\;\;
q\bar{q}\to W^+W^-  \;\;,
\]
where $l$ ($\nu$, $q$) stand for SM charged leptons (neutrinos, quarks),
as well as the $Q=1$ reactions
\[
 l\nu\to W^+Z \;\;\;,\;\;\;
q_u\bar{q}_d\to W^+Z \;\;\;,\;\;\;
q_u\bar{q}_d\to W^+\gamma \;\;\;\hbox{, and}\;\;\;
l\nu\to W^+\gamma \;\;.
\]
Taking into account that the operators $\mathcal{O}_W$,
$\mathcal{O}_B$, and $\mathcal{O}_{WWW}$ do not give rise to anomalous
triple neutral gauge boson vertices we did not consider the
$\gamma\gamma$ and $ZZ$ final states. 

Table~\ref{fermiontable} contains the unitarity violating terms for
the inelastic processes above.  As we can see, the operator
$\mathcal{O}_{WWW}$ does not contribute to the helicity amplitudes for
which $\mathcal{O}_W$ and $\mathcal{O}_B$ do due to their different
tensor structures. In order to impose unitarity constraints on these
inelastic processes we will follow the procedure described in the
previous section~\cite{Baur:1987mt}; see Eq.~(\ref{eq:unitcond2}).  We
find that strongest constraints can be imposed by using two fermion
states in the $Q=0$ ($V_aV_b=W^+W^-$) combination
\begin{eqnarray}
\ket{x1}&=&\frac{1}{\sqrt{24}}
\ket{N_f\left (-e^-_{-}e^+_{+} +{\nu_e}_{-}   {\bar{\nu_e}}_{+}
+N_c u_{-}\bar u_{+} -N_cd_{-}\bar d_{+}\right)} \;\;, \\
\ket{x2}&=&\frac{1}{\sqrt{21}}\ket{N_f\left (-e^-_{+}e^+_{-}+
N_c u_{+}\bar u_{-} -N_cd_{+}\bar d_{-}\right)} \;\;,
\end{eqnarray}
where $N_f=3$ is the number of generations and $N_C=3$ the number of
colours.  They yield the bounds
\begin{eqnarray}
&&
\frac{1}{24}\left[\left|6 \frac{g^4}{8\pi} 
\frac{f_{WWW}}{\Lambda^2} s\right |^2+
\left|1.41\frac{g^2}{8\pi} \frac{f_{W}}
{\Lambda^2} s\right|^2\right] \leq 1
\;\;\;  \Rightarrow \;\;\; 
\left|\frac{f_{WWW}}{\Lambda^2} s\right| \le 122 \; , 
\;\;\;  
\left|\frac{f_{W}}{\Lambda^2} s\right| \le 211 \;\;,
\label{eq:ffbounds}\\
&&
\frac{1}{21}\left| \sqrt{2} \frac{s_w^2}{c_w^2} \frac{g^2}{8\pi} 
\frac{f_{B}}
{\Lambda^2} s\right|^2 =
\left|0.053\frac{g^2}{8\pi} 
\frac{f_{B}}
{\Lambda^2} s\right|^2 
\leq 1
\;\;\;  \Rightarrow \;\;\; 
\left|\frac{f_{B}}{\Lambda^2}s\right| \le 664 
\nonumber
\end{eqnarray}
respectively. 

Without further information on the parameters $f_i s/\Lambda^2$, we
must consider the case where more than one of the parameters is
non-vanishing. Therefore, we should search for the largest allowed
value of a given parameter while varying over the others.
Technically we obtain these generalized bounds by searching in a
six-dimensional grid the widest range of the parameters which satisfy
both the elastic and inelastic partial-wave unitarity constraints.  We
get:
\begin{eqnarray}
\left|\frac{f_{\Phi2,4}}{\Lambda^2}s\right|
&\le    &       105 \;\;,      \nonumber \\     
\left|\frac{f_W}{\Lambda^2}s\right|                     &\le    &
205  \;\;,  \nonumber   \\      
\left|\frac{f_B}{\Lambda^2}s\right|                     &\le    &
640 \;\;, \nonumber     \\      
\left|\frac{f_{WW}}{\Lambda^2}s\right|                  &\le    &
200  \;\;,    \label{eq:finalbounds} \\      
\left|\frac{f_{BB}}{\Lambda^2}s\right|                  &\le    &
880  \;\;,  \nonumber   \\      
\left|\frac{f_{WWW}}{\Lambda^2}s\right|         &\le    &       85 \;\;.
\nonumber
\end{eqnarray}
It is important to stress that these results do not mean that the
largest ranges for each parameter can all simultaneously be realized
but rather they are the most conservative constraints on a given
parameter allowing for all possible cancellations with the others in
the scattering amplitudes.

Comparing the results in Eq.~(\ref{eq:finalbounds}) with those in
Eqs.~(\ref{eq:vvbounds}) and (\ref{eq:ffbounds}) we find that working
in the most general six-dimensinal space the bounds become weaker, but
not substantially.  Thus, even when allowing for all possible
cancellations between the contribution of the relevant dimension-six
operators, partial-wave unitarity still imposes constraints on their
range of validity.

We can now compare the unitarity constraints in
Eq.~(\ref{eq:finalbounds}) with the bounds on the corresponding
coefficients from the global analysis of the available data from
Tevatron and LHC Higgs results as well as from triple anomalous gauge
coupling bounds as updated from Ref.~\cite{us2}.  Mapping the allowed
ranges at 90\%CL of the six dimensional space from that analysis onto
the unitarity constraints in Eq.~(\ref{eq:finalbounds}) we find the
lowest energy for which presently allowed values of the coefficients
of operators affecting Higgs physics would lead to unitarity
violation. For the operator $O_{WWW}$ which only affects gauge boson
self-couplings we can naively estimate the bound by using the
presently allowed range on the effective parameter $\lambda_\gamma$
\cite{Hagiwara:1986vm} from the PDG~\cite{pdg},
$\lambda_\gamma=-0.022\pm 0.019$, which in the framework of
the dimension-six operators is related to the coefficient of
$O_{WWW}$ by $\lambda_\gamma=\lambda_Z=\frac{3}{2} M_W^2 g^2 \frac
{f_{WWW}}{\Lambda^2}$.  Altogether we find:
\begin{eqnarray}
-10\leq \frac{f_{\Phi,2}}{\Lambda^2}({\rm TeV}^{-2})\leq 8.5 &
\Rightarrow  & \sqrt{s}\leq 3.2\;{\rm TeV} \; , \nonumber \\
-5.6\leq \frac{f_{W}}{\Lambda^2}({\rm TeV}^{-2})\leq 9.6 &
\Rightarrow  & \sqrt{s}\leq 4.6\;{\rm TeV} \; , \nonumber \\
-29\leq \frac{f_{B}}{\Lambda^2}({\rm TeV}^{-2})\leq 8.9 &
\Rightarrow  & \sqrt{s}\leq 4.7\;{\rm TeV} \; , \nonumber \\
-3.2\leq \frac{f_{WW}}{\Lambda^2}({\rm TeV}^{-2})\leq 8.2 &
\Rightarrow  & \sqrt{s}\leq 4.9\;{\rm TeV} \; , \\
-7.5\leq \frac{f_{BB}}{\Lambda^2}({\rm TeV}^{-2})\leq 5.3 &
\Rightarrow  & \sqrt{s}\leq 11\;{\rm TeV} \; ,\nonumber \\
-15\leq \frac{f_{WWW}}{\Lambda^2}({\rm TeV}^{-2})\leq 3.9 &
\Rightarrow  & \sqrt{s}\leq 2.4\;{\rm TeV} \; . \nonumber 
\end{eqnarray}

In summary, in this work we have consistently derived the partial-wave
unitarity bounds on the general space of dimension-six operators
affecting Higgs and/or electroweak gauge boson interactions from
two-to-two scattering processes including vector boson and Higgs boson
scattering channels, as well as inelastic processes $f\bar f' \to V V'$
where $f^{(\prime)}$ is a SM fermion and $V^{(\prime)}$ is an electroweak gauge boson.  We
have found that the relevant set reduces to six operators and gauge
invariance enforces that the corresponding amplitudes only diverge as
$s$ in the large $s$ limit. The most general bounds obtained in this
framework are given in Eq.~(\ref{eq:finalbounds}). They can be
translated on the maximum center-of-mass energy for which the
presently allowed range of the coefficients of the corresponding
operators from the analysis of Higgs and gauge-boson data will satisfy
partial-wave unitarity.  We find that for those operators affecting
the Higgs couplings, present 90\% constrains from global data analysis
of Higgs and electroweak data are such that unitarity is not violated if
$\sqrt{s}\leq 3.2\;{\rm TeV}$.  For the purely gauge-boson operator
$O_{WWW}$, naive translation of the present bounds from triple-gauge
boson analysis indicate that within its presently allowed 90\% range
unitarity can be violated in $f\bar f' \to V V'$ at center-of-mass
energy $\sqrt{s}\geq 2.4\;{\rm TeV}$.

\section*{Acknowledgments}

We thank J. Gonzalez-Fraile for a careful reading of the manuscript
and comments.

O.J.P.E. is supported in part by Conselho Nacional de Desenvolvimento
Cient\'{\i}fico e Tecnol\'ogico (CNPq) and by Funda\c{c}\~ao de Amparo
\`a Pesquisa do Estado de S\~ao Paulo (FAPESP); M.C.G-G and T.C are 
supported by USA-NSF grants PHY-0653342 and PHY-13-16617 and by FP7 ITN
INVISIBLES (Marie Curie Actions PITN-GA-2011-289442).  M.C.G-G also
acknowledges support  by grants 2014-SGR-104 and by 
FPA2013-46570  and consolider-ingenio 2010 program CSD-2008-0037.


\appendix

\section{Anomalous interactions}
\label{app:ano}

Here we present the anomalous interactions that are generated by the
dimension--six operators in Eqs.~(\ref{eq:www})--(\ref{eq:phi3}).  For
simplicity of discussion we make use of the unitary gauge in which the
Higgs doublet becomes:
\begin{equation}
\Phi=\frac{1}{\sqrt{2}}\begin{pmatrix}0\\v+h(x)\end{pmatrix}
\end{equation}
where $v$ is the Higgs vacuum expectation value (vev).  

We first note that $\mathcal{O}_{\phi,1}$, $\mathcal{O}_{\phi,2}$, and
$\mathcal{O}_{\phi,4}$ lead to corrections of the kinetic term for the
Higgs field, therefore, we make a field redefinition to obtain a
canonical form for the kinetic Higgs term:
\begin{equation}\label{higgsrenorm}
H=h\sqrt{1+\frac{v^2}{2\Lambda^2}(f_{\Phi,1}+2f_{\Phi,2}+f_{\Phi,4})}
\;\;,
\end{equation}
resulting, together with $\mathcal{O}_{\phi,3}$, in corrections to the
Higgs mass given by (to linear order)
\begin{equation}\label{higgsmassrenorm}
M_H^2=2\lambda v^2\left(1-\frac{v^2}{2\Lambda^2}
(f_{\Phi,1}+2f_{\Phi,2}+f_{\Phi,4}+\frac{f_{\Phi,3}}{\lambda})\right)
\;\;,
\end{equation}
where $\lambda$ is the quartic scalar coupling. 
Additionally $\mathcal{O}_{BW}$ affects $Z\gamma$ mixing 
giving corrected mass eigenstates of the form:
\begin{equation}
Z_\mu=\left[1-\frac{g^2g^{\prime2}}{2(g^2+g^{\prime2})}\frac{v^2}
{\Lambda^2}f_{BW}\right]^{-1/2}Z_\mu^{SM}
\end{equation}
\begin{equation}
A_{\mu}=\left[1+\frac{g^2g^{\prime2}}{2(g^2+g^{\prime2})}
\frac{v^2}{\Lambda^2}f_{BW}\right]^{-1/2}A_\mu^{SM}-
\left[\frac{gg^\prime(g^2-g^{\prime2})}{4(g^2+g^{\prime2})}
\frac{v^2}{\Lambda^2}f_{BW}\right]Z_\mu^{SM}
\end{equation}
where:
\begin{eqnarray}
Z_{\mu}^{SM}
=\frac{1}{\sqrt{g^2+g^{\prime2}}}(gW_\mu^3-g^\prime B_\mu) &{\rm and}
&A_\mu^{SM}=\frac{1}{\sqrt{g^2+g^{\prime2}}}(g^\prime W_\mu^3+gB_\mu)
\;\;.
\end{eqnarray}
Furthermore, the operator $\mathcal{O}_{\Phi,4}$ simultaneously
affects the W and Z boson masses, while $\mathcal{O}_{\Phi,1}$ and
$\mathcal{O}_{BW}$ only affect the Z mass. Again to linear order we
have:
\begin{equation}\label{Zmassrenorm}
M_Z^2 =
\frac{g^2+g^{\prime2}}{4}v^2\left(1+\frac{v^2}{2\Lambda^2}
\left (f_{\Phi,1}+f_{\Phi,4}-\frac{g^2g^{\prime
    2}}{g^2+g^{\prime2}}f_{BW}\right)\right),
\end{equation}
\begin{equation}\label{Wmassrenorm}
M_W^2	=	\frac{g^2}{4}v^2\left(1+\frac{v^2}{2\Lambda^2}f_{\Phi,4}\right)
\;\;.
\end{equation}
Notice that in all expressions above $v$ represents the vev of the
Higgs field at the minimum of the potential including the effect of
$O_{\Phi,3}$.

We will use in our analysis as inputs the measured values of $G_F$,
$M_Z$, and $\alpha$, the so called $Z$-scheme~\cite{DeRujula:1991se},
and for convenience we absorb the tree-level renormalization factors
mentioned in equations (\ref{Zmassrenorm}) and (\ref{Wmassrenorm})
into the measured value of $M_W$. Through the relation
$\frac{G_F}{\sqrt{2}}=\frac{g^2}{8M_W^2}$ and equations
(\ref{Zmassrenorm}) and (\ref{Wmassrenorm}), we obtain the relations:
\begin{equation}\label{vrenorm}
v=(\sqrt{2} G_F)^{-1/2}\left(1-\frac{v^2}{4\Lambda^2}f_{\Phi,4}\right)
\;\;,
\end{equation}
\begin{equation}\label{Zmassrenorm2}
M_Z^2	=	(\sqrt{2}G_F)^{-1}\frac{g^2}{4\cwsq}
\left(1+\frac{v^2}{2\Lambda^2}f_{\Phi,1}-
\frac{g^2g^{\prime2}}{2(g^2+g^{\prime2})}\frac{v^2}{\Lambda^2}f_{BW}\right) \; , 
\end{equation}
where we have introduced the tree level weak mixing angle,
$\cw\equiv g/\sqrt{g^2+g^{\prime2}}$.

The dimension-six effective operators give rise to triple Higgs-gauge
interactions, taking the following forms:
\begin{eqnarray}
\mathcal{L}_{\rm eff}^{HVV}	&=
&	g_{H\gamma\gamma}HA_{\mu\nu}A^{\mu\nu}+
g_{HZ\gamma}^{(1)}A_{\mu\nu}Z^\mu\partial^\nu H +
g_{HZ\gamma}^{(2)}HA_{\mu\nu}Z^{\mu\nu} \nonumber \\
				&
+& g_{HZZ}^{(1)}Z_{\mu\nu}Z^{\mu}\partial^\nu H 
+	g_{HZZ}^{(2)}HZ_{\mu\nu}Z^{\mu\nu}
+g_{HZZ}^{(3)}HZ_{\mu}Z^{\mu} \label{effd4lagrangian1}
\\
&+&g_{HWW}^{(1)}(W_{\mu\nu}^+W^{-\mu}\partial^\nu H+{\rm h.c.})
+g_{HWW}^{(2)}HW_{\mu\nu}^+W^{-\mu\nu}+g_{HWW}^{(3)}HW_{\mu}^+W^{-\mu} \; , 
\nonumber
\end{eqnarray}
where we have defined 
$V_{\mu\nu}=\partial_\mu V_\nu-\partial_\nu V_\mu$, for $V=A,Z,W$ and 
\begin{equation}\label{d4coefficients1}
\begin{array}{lcl}
g_{H\gamma\gamma} &=& 
-\left(\frac{g^2v\swsq}{2\Lambda^2}\right)\frac{f_{BB}+f_{WW}-f_{BW}}{2}\\
g_{HZ\gamma}^{(1)}  &=& 
\left(\frac{g^2v}{2\Lambda^2}\right)\frac{\sw(f_W-f_B)}{2\cw}\\
g_{HZ\gamma}^{(2)}  &=& 
\left(\frac{g^2v}{2\Lambda^2}\right)
\frac{\sw[2\swsq f_{BB}-2\cwsq f_{WW}+(\cwsq-\swsq)f_{BW}]}{2\cw}\\
g_{HZZ}^{(1)}   &=& 
\left(\frac{g^2v}{2\Lambda^2}\right)\frac{\cwsq f_W+\swsq f_B}{2\cwsq}\\
g_{HZZ}^{(2)}   &=& 
-\left(\frac{g^2v}{2\Lambda^2}\right)
\frac{s_{\rm W}^4f_{BB}+c_{\rm W}^4f_{WW}+\cwsq \swsq f_{BW}}{2\cwsq}\\
g_{HZZ}^{(3)}   &=& 
\left(\frac{g^2v}{4\cwsq}\right)
\left[1+\frac{v^2}{4\Lambda^2}
\left(3f_{\Phi,1}+3f_{\Phi,4}-2f_{\Phi,2}-
\frac{2g^2g^{\prime2}}{(g^2+g^{\prime2})}f_{BW}\right)\right]\\
&=& M_Z^2(\sqrt{2}G_F)^{1/2}\left[1+\frac{v^2}{4\Lambda^2}
(f_{\Phi,1}+2f_{\Phi,4}-2f_{\Phi,2})\right]\\
g_{HWW}^{(1)}   &=& 
\left(\frac{g^2v}{2\Lambda^2}\right)\frac{f_{W}}{2}\\
g_{HWW}^{(2)}   &=& 
-\left(\frac{g^2v}{2\Lambda^2}\right)f_{WW}\\
g_{HWW}^{(3)}   &=& 
\left(\frac{g^2v}{2}\right)
\left[1+\frac{v^2}{4\Lambda^2}(3f_{\Phi4}-f_{\Phi,1}-2f_{\Phi,2})\right]\\
&=& 2M_W^2(\sqrt{2}G_F)^{1/2}\left[1+\frac{v^2}{4\Lambda^2}
(2f_{\Phi,4}-f_{\Phi,1}-2f_{\Phi,2})\right]
\end{array}
\end{equation}
Quartic vertices involving Higgs and gauge bosons read:
\begin{eqnarray}
\mathcal{L}_{\rm eff}^{HHV_1V_2}	&=&	
g_{HHWW}^{(1)}H^2 W_{\mu\nu}^+W^{-\mu\nu}+
g_{HHWW}^{(2)}H(\partial_\nu H)(W_{\mu}^-W^{+\mu\nu}+{\rm h.c.}) 
\nonumber\\						
&& +g_{HHWW}^{(3)}H^2W_\mu^+W^{-\mu} 
\nonumber
\\
&+&	g_{HHZZ}^{(1)}H^2Z_{\mu\nu}Z^{\mu\nu}+g_{HHZZ}^{(2)}HZ_\nu
(\partial_\mu H)Z^{\mu\nu}+g_{HHZZ}^{(3)}H^2Z_\mu Z^\mu \label{effd4lagrangian2}\\
&+& 	g_{HHZA}^{(1)}H(\partial_\mu H)Z_\nu A^{\mu\nu}+
g_{HHZA}^{(2)}H^2A_{\mu\nu}Z^{\mu\nu} \nonumber\\
&+& g_{HHAA}^{(1)}H^2A_{\mu\nu}A^{\mu\nu} \; ,  \nonumber 
\end{eqnarray}
with 
\begin{equation}\label{d4coefficients2}
\begin{array}{lcl}
g_{HHWW}^{(1)}  &=& -\frac{g^2}{4\Lambda^2}f_{WW}\\
g_{HHWW}^{(2)}  &=& \frac{g^2}{4\Lambda^2}f_W\\
g_{HHWW}^{(3)}  &=& \frac{g^2}{4}
\left[1+\frac{v^2}{2\Lambda^2}( 5f_{\Phi,4}-f_{\Phi,1}-2f_{\Phi,2})\right]\\
&=&{ M_W^2 \sqrt{2}G_F\left[1+\frac{v^2}{2\Lambda^2}( 5f_{\Phi,4}-f_{\Phi,1}-2f_{\Phi,2})\right]}\\
g_{HHZZ}^{(1)}   &=& 
-\frac{g^2}{8\cwsq\Lambda^2}(c_{\rm W}^4f_{WW}+s_{\rm W}^4f_{BB}+\cwsq \swsq f_{BW})\\
g_{HHZZ}^{(2)}   &=& 
-\frac{g^2}{4\cwsq\Lambda^2}(\cwsq f_W+\swsq f_B)\\
g_{HHZZ}^{(3)}   &=& 
\frac{g^2}{8\cwsq}
\left[1+\frac{v^2}{2\Lambda^2}(5f_{\Phi,1}+5f_{\Phi,4}-2f_{\Phi,2} 
{-\frac{g^2g^{\prime2}}{(g^2+g^{\prime2})}f_{BW}})\right]\\
&=&
{M_Z^2 \sqrt{2}G_F
\left[1+\frac{v^2}{2\Lambda^2}(4 f_{\Phi,1}+5f_{\Phi,4}-2f_{\Phi,2}) \right]}
\\
g_{HHZA}^{(1)}   &=& -\frac{g^2\sw}{4\cw\Lambda^2}(f_W-f_B)\\
g_{HHZA}^{(2)}   &=& 
-\frac{g^2\sw}{4\cw\Lambda^2}(\cwsq f_{WW}-\swsq f_{BB}
-\frac{1}{2}(\cwsq-\swsq)f_{BW})\\
g_{HHAA}^{(1)}  &=& 
-\frac{g^2\swsq}{8\Lambda^2}(f_{WW}+f_{BB}-f_{BW})

\end{array}
\end{equation}
and 
\begin{eqnarray}
\mathcal{L}_{\rm eff}^{HV_1V_2V_3}	&=&	
g_{HZWW}^{(1)}H(W_{\mu}^-W_{\nu}^+ -
{\rm h.c.})Z^{\mu\nu} 
+ g_{HZWW}^{(2)}HZ_\mu (W_\nu^+W^{-\mu\nu}
-
{\rm h.c.})
+g_{HZWW}^{(3)}(\partial_\mu H)Z_\nu(W^{-\mu}W^{+\nu}
-{\rm h.c.}) 
\nonumber \\
&+&	g_{HAWW}^{(1)}H(W_{\mu}^-W_{\nu}^+
-{\rm h.c.})A_{\mu\nu}+
g_{HAWW}^{(2)}HA_\nu(W_\mu^{+\nu}W^{-\mu}
-{\rm h.c.})
\label{effd4lagrangian3}\\
&+& g_{HAWW}^{(3)}(\partial_\mu H)A_\nu(W^{-\mu}W^{+\nu}
-{\rm h.c.})
\; ,  \nonumber
\end{eqnarray}
with
\begin{equation}\label{d4coefficients3}
\begin{array}{lcl}
g_{HZWW}^{(1)}  &=& 
\frac{ig^3 v}{8\cw\Lambda^2}(\cwsq f_W - \swsq f_B+4\cwsq f_{WW}
+2\swsq f_{BW})\\
g_{HZWW}^{(2)}  &=& 
-\frac{ig^3 v}{4\cw\Lambda^2}(f_W+4\cwsq f_{WW})\\
g_{HZWW}^{(3)}  &=& 
\frac{ig^3v}{4\cw\Lambda^2}\swsq f_W\\
g_{HAWW}^{(1)}  &=& 
\frac{ig^3v\sw}{8\Lambda^2}(
f_W+f_B+4f_{WW}-2f_{BW})\\
g_{HAWW}^{(2)}  &=& 
-\frac{ig^3\sw v}{\Lambda^2} f_{WW}\\
g_{HAWW}^{(3)}  &=& -\frac{ig^3v\sw}{4\Lambda^2}f_W
\end{array}
\end{equation}
Triple gauge boson couplings are:
\begin{eqnarray}
\mathcal{L}_{\rm eff}^{WWV} 	&=&	
g_{WWZ}^{(1)} (W_{\nu}^+W_{\mu}^- -{\rm h.c.})Z^{\mu\nu} 
+g_{WWZ}^{(2)}(W_{\mu\nu}^{+}W^{-\mu}Z^\nu-{\rm h.c.})
+g_{WWZ}^{(3)} (W_{\mu\nu}^+W_{\rho}^{-\nu}-{\rm h.c.})Z^{\rho\mu}\nonumber\\
						&
+&	g_{WWA}^{(1)} (W_{\nu}^+W_{\mu}^- -{\rm h.c.})A^{\mu\nu}
+g_{WWA}^{(2)}(W_{\mu\nu}^+W_{\rho}^{-\nu}-{\rm h.c.})A^{\rho\mu} \; , 
\label{effd4lagrangian4}
\end{eqnarray}
where 
\begin{equation}\label{d4coefficients4}
\begin{array}{lcl}
g_{WWZ}^{(1)}   &=& 
\frac{i g^3v^2\cw}{16\Lambda^2}(f_W+\frac{\swsq}{\cwsq}f_{B}+
{
\frac{4\swsq}{\ctw}} f_{BW}-\frac{2\swsq}{e^2\ctw}f_{\Phi,1})
{\equiv \frac{i g\cw}{2} \Delta \kappa_Z}
\\
g_{WWZ}^{(2)}   &=& 
-\frac{i g^3 v^2}{8\cw\Lambda^2} (f_W
+\frac{2\swsq}{\ctw}f_{BW}-\frac{\stw^2}{2e^2\ctw}f_{\Phi,1})
{\equiv {-i g\cw} \Delta g_Z}
\\
g_{WWZ}^{(3)}   &=& 
-\frac{3ig^3\cw\Lambda^2}{2}f_{WWW} {\equiv
\frac{-i g\cw}{M_W^2}\lambda_Z}
\\
g_{WWA}^{(1)}   &=& 
\frac{i g^3 v^2 \sw}{16 \Lambda^2}(f_W+f_B-2f_{BW})
{\equiv\frac{i g\sw}{2}\Delta\kappa_\gamma}
\\
g_{WWA}^{(2)}   &=& 
-\frac{3ig^3 \sw}{2 \Lambda^2}f_{WWW}
{\equiv\frac{-i g\sw}{M_W^2}\lambda_\gamma}
\end{array}
\end{equation}
where we have defined $\ctw=\cos(2\theta_w)$ and $\stw=\sin(2\theta_w)$. 

Quartic gauge boson vertices read:
\begin{eqnarray}
\mathcal{L}_{\rm eff}^{WWV_1V_2}        &=&
g_{WWWW}^{(1)} W_{\mu}^{-}W_{\nu}^{+}(W^{-\mu}W^{+\nu}-{\rm h.c.})
+g_{WWWW}^{(2)}W_{\mu\nu}^+W^{-\nu\rho}(W^{+\mu}W^{-}_\rho-W^{+}_\rho W^{-\mu})
\nonumber\\
&+&     g_{WWZZ}^{(1)}Z_\mu Z^\mu W_\nu^+W_\nu^-
+g_{WWZZ}^{(2)}Z_\mu Z_\nu(W_\nu^+W_\mu^- +
{\rm h.c.})
+g_{WWZZ}^{(3)}\left(W^+_{\mu\nu}Z^{\phantom{\nu}\mu}_\rho(Z^\nu W^{-\rho}-
Z^{\rho}W^{-\nu})+{\rm h.c.}\right)
\nonumber\\
&+&     g_{WWAA}^{(3)}
\left(W^+_{\mu\nu}A^{\phantom{\nu}\mu}_\rho
(A^\nu W^{-\rho}-A^{\rho}W^{-\nu})+{\rm h.c.}\right)
\nonumber\\
&+&     g_{WWZA}^{(1)}W_\mu^-W^{+\mu}Z_\mu A^\mu
+ g_{WWZA}^{(2)}(W_\nu^-W_\mu^++{\rm h.c.})A^\nu Z^\mu
\label{effd4lagrangian5}\\
&+&g_{WWZA}^{(3)}\left(W^+_{\mu\nu}Z^{\phantom{\nu}\mu}_\rho
(A^\nu W^{-\rho}- A^{\rho}W^{-\nu})+W^+_{\mu\nu}A^{\phantom{\nu}\mu}_\rho
(Z^\nu W^{-\rho}-Z^{\rho} W^{-\nu})+{\rm h.c.}\right)
\nonumber
\end{eqnarray}
with
\begin{equation}\label{d4coefficients5}
\begin{array}{lcl}
g_{WWWW}^{(1)}  &=&
{\frac{e^2}{2 \swsq}}+\frac{g^4v^2}{8\Lambda^2}(f_W+2\frac{\swsq}{\ctw}f_{BW}-\frac{
\stw^2}{{2}\ctw e^2}f_{\Phi,1}) \\
g_{WWWW}^{(2)}  &=& 
\frac{-3g^4}{2\Lambda^2}f_{WWW}\\
g_{WWZZ}^{(1)}  &=& -e^2 \frac{\cwsq}{\swsq}
-\frac{g^4v^2\Lambda^2}{4\cwsq}(\cwsq f_W+
\frac{2\stw^2}{\ctw}f_{BW}
-\frac{\stw^2\cwsq}{2e^2\ctw}f_{\Phi,1})\\
g_{WWZZ}^{(2)}  &=& \frac{e^2 \cwsq}{2\swsq}
+\frac{g^4v^2\Lambda^2}{8\cwsq}(\cwsq f_W+\frac{\stw^2}{2\ctw}f_{BW}-\frac{\stw^2\cwsq}{2e^2\ctw}f_{\Phi,1})\\
g_{WWZZ}^{(3)}  &=& 
\frac{-3g^4v^2\cwsq}{2\Lambda^2}f_{WWW}\\
g_{WWAA}^{(3)}  &=& 
-\frac{3g^4v^2\swsq}{2\Lambda^2}f_{WWW}\\
g_{WWZA}^{(1)}  &=& -e^2 
-\frac{g^4v^2\sw}{4\cw\Lambda^2}(f_W+2\frac{
\swsq}{\ctw}f_{BW}-\frac{\stw^2}{2\ctw e^2}f_{\Phi,1})\\
g_{WWZA}^{(2)}  &=& \frac{e^2}{2}
+\frac{g^4v^2\sw}{8\cw\Lambda^2}(f_W+2\frac{\swsq}{\ctw}f_{BW}-\frac{\stw^2}{2\ctw e^2}f_{\Phi,1})\\
g_{WWZA}^{(3)}  &=& 
\frac{-3g^4\sw\cw}{2\Lambda^2}f_{WWW}
\end{array}
\end{equation}
Finally Higgs self interactions take the form:
\begin{eqnarray}
\mathcal{L}_{\rm eff}^{HHH}&=&
g_{HHH}^{(1)}H^3
+ g_{HHH}^{(2)}H(\partial_\mu H)(\partial^\mu H) \; , \\
\\
\mathcal{L}_{\rm eff}^{HHHH}&=&
g_{HHHH}^{(1)}H^4
+g_{HHHH}^{(2)}H^2(\partial_\mu H)(\partial^\mu H) \; , 
\end{eqnarray}
where 
\begin{equation}\label{d4coefficients6}
\begin{array}{lcl}
g_{HHH}^{(1)}&=&
-\lambda v+\frac{v^3}{\Lambda^2}(\frac{3\lambda}{4}f_{\Phi,1}+
{\frac{5}{6}}f_{\Phi,3}+\frac{3\lambda}{2} f_{\Phi,2}+\frac{3\lambda}{4}f_{\Phi,4})\\
&=&-\frac{M_H^2}{2}(\sqrt{2}G_F)^{1/2}\left[1
-\frac{v^2}{4\Lambda^2}(f_{\Phi,1}+2f_{\Phi,2}
{\frac{4}{3\lambda}}f_{\Phi,3})\right]\\
g_{HHH}^{(2)}&=&
\frac{v}{\Lambda^2}(\frac{1}{2}f_{\Phi,1}+f_{\Phi,2}+\frac{1}{2}f_{\Phi,4})\\
g_{HHHH}^{(1)}&=&
-\frac{\lambda}{4}+\frac{v^2}{4\Lambda^2}(\lambda f_{\Phi,1}+
{\frac{5}{2}f_{\Phi,3}}+2\lambda f_{\Phi,2}+\lambda f_{\Phi,4})\\
&=&-\frac{M_H^2}{8}(\sqrt{2}G_F)\left[1+\frac{v^2}{2\Lambda^2}
(f_{\Phi,1}+{\frac{4}{\lambda}}f_{\Phi,3}+f_{\Phi,2})\right]\\
g_{HHHH}^{(2)}&=&
\frac{1}{4\Lambda^2}(f_{\Phi,1}+ 2 f_{\Phi,2}+f_{\Phi,4}) 
\end{array}
\end{equation}

\section{Helicity Amplitudes}
\label{app:tables}
We present here the list of unitarity violating amplitudes for
all the $2\rightarrow 2$ scattering processes considered in the evaluation
of the unitarity constraints. 
\begin{table}[h]
\begin{tabular}{|l||c|}
\hline
 & $(\times \frac{f_{\Phi,2,4}}{\Lambda^2} \times s)$  \\ \hline
$W^+W^+ \rightarrow W^+W^+$ 				&$	-1$ \\ 
\hline
$W^+Z \rightarrow W^+Z$						
&$	-\frac{1}{2}X		$ \\ 
$W^+ H \rightarrow W^+ H$ 		&$	-\frac{1}{2}X	$\\ 
\hline
$W^+W^- \rightarrow W^+W^-$ 		&$	\frac{1}{2}Y	$\\ 
$W^+W^- \rightarrow Z Z$ 		&$	1		$\\ 
$W^+W^- \rightarrow H H$ 		&$	-1		$\\ 
$Z Z \rightarrow  H H $ 		&$	-1		$\\
$Z H \rightarrow  Z H $ 		&$	-\frac{1}{2}X	$\\   
\hline
\end{tabular}
\caption{Unitarity violating (growing as $s$) terms of the scattering
  amplitudes $\mathcal{M}({V_1}_{\lambda_1}{V_2 }_{\lambda_2} \to
  {V_3}_{\lambda_3}{V_4}_{\lambda_4})$ for longitudinal gauge bosons
  generated by the operators ${\cal O}_{\Phi,2}$ and ${\cal
    O}_{\Phi,4}$ where $X=1-\cos\theta$ and $Y=1+\cos\theta$.  The
  overall factor extracted from all amplitudes is given at the top of
  the table.
\label{tab:fphis}}
\end{table}
\begin{table}[h]
\begin{tabular}{|c||c|c|c|c|c|c|c|}
\hline
 & \multicolumn{7}{c|}{$(\times e^2 \frac{f_{W}}{\Lambda^2}\times s)$}  
\\ \hline &
$0000$ & $00++$ & $0+0-$ & $0+-0$ & $+00-$ & $+0-0$ & $++00$ \\ \hline
$W^+W^+ \rightarrow W^+W^+$& 
	$	-\frac{3}{4\swsq}	$&$	0	$&$	\frac{1}{8\swsq}X	$&$	-\frac{1}{8\swsq}Y	$&$	-\frac{1}{8\swsq}Y	$&$	\frac{1}{8\swsq}X	$&$	0	$  
\\ 
\hline
$W^+Z \rightarrow W^+Z$& 
	$	-\frac{3}{8\swsq}X	$&$	-\frac{1}{8\cw}	$&$	\frac{\cwsq-\swsq}{8\swsq}X	$&$	-\frac{1}{16\cw}Y	$&$	-\frac{1}{16\cw}Y	$&$	\frac{1}{8\cwsq}X	$&$	-\frac{1}{8\cw}	$
\\ 
$W^+\gamma \rightarrow W^+\gamma$& 
	$	\na	$&$	\na	$&$	\frac{1}{4}X	$&$	\na	$&$	\na	$&$	\na	$&$	\na	$
\\ 
$W^+Z \rightarrow W^+\gamma$& 
	$	\na	$&$	\frac{1}{8\sw}	$&$	\frac{(3\cwsq-\swsq)}{16\cw\sw}X	$&$	\na	$&$	\frac{1}{16\sw}Y	$&$	\na	$&$	\na	$
\\ 
$W^+ Z \rightarrow W^+ H$& 
	$	0				$&$	\na	$&$	\na	$&$	-\frac{1}{16\cw}Y	$&$	\na	$&$	0	$&$	\frac{1}{8\cw}	$
\\ 
$W^+ \gamma \rightarrow W^+ H$ &
	$	\na	$&$	\na	$&$	\na	$&$	\frac{1}{16\sw}Y	$&$	\na	$&$	\na	$&$	-\frac{1}{8\sw}	$
\\ 
$W^+ H \rightarrow W^+ H$ &
	$	-\frac{3}{8\swsq}X		$&$	\na	$&$	\na	$&$	\na	$&$	\na	$&$	\frac{1}{8\swsq}X	$&$	\na	$
\\ 
 \hline
$W^+W^- \rightarrow W^+W^-$ &   
	$	\frac{3}{8\swsq}Y	$&$	-\frac{1}{4\swsq}	$&$	\frac{1}{8\swsq}X	$&$	0	$&$	0	$&$	\frac{1}{8\swsq}X	$&$	-\frac{1}{4\swsq}	$
\\ 
$W^+W^- \rightarrow Z Z$ &
	$	\frac{3}{4\swsq}	$&$	\frac{\swsq-\cwsq}{4\swsq}		$&$	\frac{1}{16\cw}X	$&$	-\frac{1}{16\cw}Y	$&$	-\frac{1}{16\cw}Y	$&$	\frac{1}{16\cw}X	$&$	-\frac{1}{4\swsq}	$
\\ 
$W^+W^- \rightarrow \gamma\gamma$ & 
	$	\na	$&$	-\frac{1}{2}	$&$	\na	$&$	\na	$&$	\na	$&$	\na	$&$	\na	$
\\ 
$W^+W^- \rightarrow Z \gamma$ &
	$	\na	$&$	\frac{1-4\cwsq}{8\cw\sw}	$&$	-\frac{1}{16\sw}X	$&$	\na	$&$	\frac{1}{16\sw}Y	$&$	\na	$&$	\na	$
\\  
$W^+W^- \rightarrow Z H$ &
	$	0	$&$	\na	$&$	\na	$&$	-\frac{1}{16\cw}Y	$&$	\na	$&$	-\frac{1}{16\cw}X	$&$	0	$
\\ 
$W^+W^- \rightarrow \gamma H$ & 
	$	\na	$&$	0	$&$	\na	$&$	\frac{1}{16\sw}Y	$&$	\na	$&$	\frac{1}{16\sw}X	$&$	\na	$
\\   
$W^+W^- \rightarrow H H$ &
	$	-\frac{3}{4\swsq}	$&$	\na	$&$	\na	$&$	\na	$&$	\na	$&$	\na	$&$	\frac{1}{4\swsq}	$
\\ 
$Z Z \rightarrow  Z Z$ &
	$	0	$&$	-\frac{1}{4\swsq}	$&$	\frac{1}{8\swsq}X	$&$	-\frac{1}{8\swsq}Y	$&$	-\frac{1}{8\swsq}Y	$&$	\frac{1}{8\swsq}X	$&$	-\frac{1}{4\swsq}	$
\\ 
$Z Z \rightarrow  Z \gamma $ &
	$	\na	$&$	-\frac{1}{8\cw\sw}	$&$	\frac{1}{16\cw\sw}X	$&$	\na	$&$	-\frac{1}{16\cw\sw}Y	$&$	\na	$&$	\na	$
\\ 
$Z Z \rightarrow  H H $ &
	$	-\frac{3}{4\swsq}	$&$	\na	$&$	\na	$&$	\na	$&$	\na	$&$	\na	$&$	\frac{1}{4\swsq}	$
\\ 
$Z \gamma \rightarrow  Z Z$ &
	$	\na	$&$	\na	$&$	\frac{1}{16\cw\sw}X	$&$	-\frac{1}{16\cw\sw}Y	$&$	\na	$&$	\na	$&$	-\frac{1}{8\sw\cw}	$
\\ 
$Z \gamma \rightarrow  H H $ &
	$	\na	$&$	\na	$&$	\na	$&$	\na	$&$	\na	$&$	\na	$&$	\frac{1}{8\sw\cw}	$
\\ 
$Z H \rightarrow  Z H $ &
	$	-\frac{3}{8\swsq}X	$&$	\na	$&$	\na	$&$	\na	$&$	\na	$&$	\frac{1}{8\swsq}X	$&$	\na	$
\\  
$Z H \rightarrow  \gamma H $ &
	$	\na	$&$	\na	$&$	\na	$&$	\na	$&$	\na	$&$	\frac{1}{16\cw\sw}X	$&$	\na	$
\\
\hline
\end{tabular}
\caption{ Unitarity violating (growing as $s$) terms of the scattering
  amplitudes $\mathcal{M}({V_1}_{\lambda_1}{V_2 }_{\lambda_2} \to
  {V_3}_{\lambda_3}{V_4}_{\lambda_4})$ for gauge bosons with the
  helicities $\lambda_1\lambda_2\lambda_2\lambda_4$ listed on top of
  each column, generated by the operator ${\cal O}_{W}$.
  $X=1-\cos\theta$ and $Y=1+\cos\theta$.  The overall factor extracted
  from all amplitudes is given on the top of the table. An entry
  marked as 0 means that there is no $s$ growth for the amplitude,
  while we denote as $\na$ an amplitude that does not exist.
\label{tab:fw}}
\end{table}
\begin{table}[n]
\begin{tabular}{|c||c|c|c|c|c|c|c|}
\hline
& \multicolumn{7}{c|}{$(\times e^2 \frac{f_{B}}{\Lambda^2})\times s$}   
\\ \hline &
$0000$ & $00++$ & $0+0-$ & $0+-0$ & $+00-$ & $+0-0$ & $++00$ \\ \hline
$W^+W^+ \rightarrow W^+W^+$& 
   	$	-\frac{3}{4\cwsq}	$&$	0	$&$	0	$&$	0	$&$	0	$&$	0	$ &$	0	$
\\ 
\hline
$W^+Z \rightarrow W^+Z$& 
   	$	0	$&$	-\frac{1}{8\cw}	$&$	\frac{\swsq-\cwsq}{8\cwsq}X	$&$	-\frac{1}{16\cw}Y	$&$	-\frac{1}{16\cw}Y	$&$	0	$&$	-\frac{1}{8\cw}	$
\\ 
$W^+\gamma \rightarrow W^+\gamma$& 
   	$	\na	$&$	\na	$&$	\frac{1}{4}X	$&$	\na	$&$	\na	$&$	\na	$&$	\na	$
\\ 
$W^+Z \rightarrow W^+\gamma$& 
   	$	\na	$&$	\frac{1}{8\sw}	$&$	\frac{\cwsq-3\swsq}{16\sw\cw}X	$&$	\na	$&$	\frac{1}{16\sw}Y	$&$	\na	$&$	\na	$
\\ 
$W^+ Z \rightarrow W^+ H$& 
   	$	-\frac{2+Y}{8\cwsq}	$&$	\na	$&$	\na	$&$	-\frac{1}{16\cw}Y	$&$	\na	$&$	0	$&$	\frac{1}{8\cw}	$
\\ 
$W^+ \gamma \rightarrow W^+ H$ &
   	$	\na	$&$	\na	$&$	\na	$&$	\frac{1}{16\sw}Y	$&$	\na	$&$	\na	$&$	-\frac{1}{8\sw}	$
\\ 
 \hline
$W^+W^- \rightarrow W^+W^-$ &   
   	$	\frac{3}{8\cwsq}Y	$&$	0	$&$	0	$&$	0	$&$	0	$&$	0	$&$	0	$
\\ 
$W^+W^- \rightarrow Z Z$ &
   	$	0			$&$	\frac{\cwsq-\swsq}{4\cwsq}	$&$	\frac{1}{16\cw}X	$&$	-\frac{1}{16\cw}Y	$&$	-\frac{1}{16\cw}Y	$&$	\frac{1}{16\cw}X	$&$	0	$
\\ 
$W^+W^- \rightarrow \gamma\gamma$ & 
   	$	\na	$&$	-\frac{1}{2}		$&$	\na	$&$	\na	$&$	\na	$&$	\na	$&$	\na	$
\\ 
$W^+W^- \rightarrow Z \gamma$ &
   	$	\na	$&$	\frac{3-4\cwsq}{8\cw\sw}	$&$	-\frac{1}{16\sw}X	$&$	\na	$&$	\frac{1}{16\sw}Y	$&$	\na	$&$	\na	$   
\\  
$W^+W^- \rightarrow Z H$ &
   	$	\frac{1-Y}{4\cwsq}	$&$	\na	$&$	\na	$&$	-\frac{1}{16\cw}Y	$&$	\na	$&$	-\frac{1}{16\cw}X	$&$	0	$
\\ 
$W^+W^- \rightarrow \gamma H$ & 
   	$	\na	$&$	0	$&$	\na	$&$	\frac{1}{16\sw}Y	$&$	\na	$&$	\frac{1}{16\sw}X	$&$	\na	$   
\\   
$Z Z \rightarrow  Z Z$ &
   	$	0	$&$	-\frac{1}{4\cwsq}	$&$	\frac{1}{8\cwsq}X	$&$	-\frac{1}{8\cwsq}Y	$&$	-\frac{1}{8\cwsq}Y	$&$	\frac{1}{8\cwsq}X	$&$	-\frac{1}{4\cwsq}	$   
\\ 
$Z Z \rightarrow  Z \gamma $ &
   	$	\na	$&$	\frac{1}{8\cw\sw}	$&$	-\frac{1}{16\cw\sw}X	$&$	\na	$&$	\frac{1}{16\cw\sw}Y	$&$	\na	$&$	\na	$   
\\ 
$Z Z \rightarrow  H H $ &
   	$	-\frac{3}{4\cwsq}	$&$	\na	$&$	\na	$&$	\na	$&$	\na	$&$	\na	$&$	\frac{1}{4\cwsq}	$   
\\ 
$Z \gamma \rightarrow  Z Z$ &
   	$	\na	$&$	\na	$&$	-\frac{1}{16\cw\sw}X	$&$	\frac{1}{16\cw\sw}Y	$&$	\na	$&$	\na	$&$	\frac{1}{8\sw\cw}	$   
\\ 
$Z \gamma \rightarrow  H H $ &
   	$	\na	$&$	\na	$&$	\na	$&$	\na	$&$	\na	$&$	\na	$&$	-\frac{1}{8\sw\cw}	$   
\\ 
$Z H \rightarrow  Z H $ &
   	$	-\frac{3}{8\cwsq}X	$&$	\na	$&$	\na	$&$	\na	$&$	\na	$&$	\frac{1}{8\cwsq}X	$&$	\na	$   
\\  
$Z H \rightarrow  \gamma H $ &
   	$	\na	$&$	\na	$&$	\na	$&$	\na	$&$	\na	$&$	-\frac{1}{16\cw\sw}X	$&$	\na	$  
\\
\hline
\end{tabular}
\caption{Same as Table~\ref{tab:fw} for the operator ${\cal O}_{B}$.
\label{tab:fb}}
\end{table}
\begin{table}[n]
\begin{tabular}{|c|| c| c| c| c| c| c || c| c| c| c| c| c|}
\hline
& \multicolumn{6}{c|}{$(\times e^2 \frac{f_{WW}}{\Lambda^2}\times s)$}  
& \multicolumn{6}{c|}{$(\times e^2 \frac{f_{BB}}{\Lambda^2})\times s$}   
\\ \hline &
	$00++$ 		& $0+0-$ 		& $0+-0$ 		& $+00-$		 & $+0-0$ 		& $++00$ &
	$00++$ 		& $0+0-$ 		& $0+-0$ 		& $+00-$ 		& $+0-0$ 		& $++00$ \\ \hline
$W^+W^+ \rightarrow W^+W^+$& 
	$	0	$&$	-\frac{1}{4\swsq}X	$&$	\frac{1}{4\swsq}Y	$&$	\frac{1}{4\swsq}Y	$&$	-\frac{1}{4\swsq}X	$&$	0	$&
   	$	0	$&$	0				$&$	0				$&$	0				$&$	0				$&$	0	$  
\\ 
\hline
$W^+Z \rightarrow W^+Z$& 
	$	0	$&$	-\frac{\cwsq}{4\swsq}X	$&$	0	$&$	0	$&$	-\frac{1}{4\swsq}X	$&$	0	$&
   	$	0	$&$	-\frac{\swsq}{4\cwsq}X	$&$	0	$&$	0	$&$	0				$&$	0	$   
\\ 
$W^+\gamma \rightarrow W^+\gamma$& 
	$	\na	$&$	-\frac{1}{4}X	$&$	\na	$&$	\na	$&$	\na	$&$	\na	$&
   	$	\na	$&$	-\frac{1}{4}X	$&$	\na	$&$	\na	$&$	\na	$&$	\na	$   
\\ 
$W^+Z \rightarrow W^+\gamma$& 
	$	0	$&$	-\frac{\cw}{4\sw}X	$&$	\na	$&$	0	$&$	\na	$&$	\na	$&
   	$	0	$&$	\frac{\sw}{4\cw}X	$&$	\na	$&$	0	$&$	\na	$&$	\na	$   
\\ 
$W^+ H \rightarrow W^+ H$ &
	$	\na	$&$	\na	$&$	\na	$&$	\na	$&$	-\frac{1}{4\swsq}X	$&$	\na	$&
   	$	\na	$&$	\na	$&$	\na	$&$	\na	$&$	0				$&$	\na	$   
\\ 
 \hline
$W^+W^- \rightarrow W^+W^-$ &   
	$	\frac{1}{2\swsq}	$&$	-\frac{1}{4\swsq}X	$&$	0	$&$	0	$&$	-\frac{1}{4\swsq}X	$&$	\frac{1}{2\swsq}	$&
   	$	0			$&$	0				$&$	0	$&$	0	$&$	0				$&$	0			$   
\\ 
$W^+W^- \rightarrow Z Z$ &
	$	\frac{\cwsq}{2\swsq}	$&$	0	$&$	0	$&$	0	$&$	0	$&$	\frac{1}{2\swsq}	$&
   	$	\frac{\swsq}{4\cwsq}	$&$	0	$&$	0	$&$	0	$&$	0	$&$	0			$   
\\ 
$W^+W^- \rightarrow \gamma\gamma$ & 
	$	\frac{1}{2}	$&$	\na	$&$	\na	$&$	\na	$&$	\na	$&$	\na	$&
   	$	\frac{1}{2}	$&$	\na	$&$	\na	$&$	\na	$&$	\na	$&$	\na	$   
\\ 
$W^+W^- \rightarrow Z \gamma$ &
	$	\frac{\cw}{2\sw}		$&$	0	$&$	\na	$&$	0	$&$	\na	$&$	\na	$&
   	$	-\frac{\sw}{2\cw}	$&$	0	$&$	\na	$&$	0	$&$	\na	$&$	\na	$   
\\  
$W^+W^- \rightarrow H H$ &
	$	\na	$&$	\na	$&$	\na	$&$	\na	$&$	\na	$&$	-\frac{1}{2\swsq}	$&
   	$	\na	$&$	\na	$&$	\na	$&$	\na	$&$	\na	$&$	0				$   
\\ 
$Z Z \rightarrow  Z Z$ &
	$	\frac{\cwsq}{2\swsq}	$&$	-\frac{\cwsq}{4\swsq}X	$&$	\frac{\cwsq}{4\swsq}Y	$&$	\frac{\cwsq}{4\swsq}Y	$&$	-\frac{\cwsq}{4\swsq}X	$&$	\frac{\cwsq}{2\swsq}	$&
   	$	\frac{\swsq}{2\cwsq}	$&$	-\frac{\swsq}{4\cwsq}X	$&$	\frac{\swsq}{4\cwsq}Y	$&$	\frac{\swsq}{4\cwsq}Y	$&$	-\frac{\swsq}{4\cwsq}X	$&$	\frac{\swsq}{2\cwsq}	$   
\\ 
$Z Z \rightarrow  \gamma \gamma $ & 
	$	\frac{1}{2}	$&$	\na	$&$	\na	$&$	\na	$&$	\na	$&$	\na	$&
   	$	\frac{1}{2}	$&$	\na	$&$	\na	$&$	\na	$&$	\na	$&$	\na	$   
\\ 
$Z Z \rightarrow  Z \gamma $ &
	$	\frac{\cw}{2\sw}		$&$	-\frac{\cw}{4\sw}X	$&$	\na	$&$	\frac{\cw}{4\sw}Y	$&$	\na	$&$	\na	$&
   	$	-\frac{\sw}{2\cw}	$&$	\frac{\sw}{4\cw}X	$&$	\na	$&$	-\frac{\sw}{4\cw}Y	$&$	\na	$&$	\na	$   
\\ 
$Z Z \rightarrow  H H $ &
	$	\na	$&$	\na	$&$	\na	$&$	\na	$&$	\na	$&$	-\frac{\cwsq}{2\swsq}	$&    
   	$	\na	$&$	\na	$&$	\na	$&$	\na	$&$	\na	$&$	-\frac{\swsq}{2\cwsq}	$   
\\ 
$Z \gamma \rightarrow  Z Z$ &
	$	\na	$&$	-\frac{\cw}{4\sw}X	$&$	\frac{\cw}{4\sw}Y	$&$	\na	$&$	\na	$&$	\frac{\cw}{2\sw}	$&    
   	$	\na	$&$	\frac{\sw}{4\cw}X	$&$	-\frac{\sw}{4\cw}Y	$&$	\na	$&$	\na	$&$	-\frac{\sw}{2\cw}	$   
\\ 
$Z \gamma \rightarrow  Z \gamma $ &
	$	\na	$&$	-\frac{1}{4}X	$&$	\na	$&$	\na	$&$	\na	$&$	\na	$&    
   	$	\na	$&$	-\frac{1}{4}X	$&$	\na	$&$	\na	$&$	\na	$&$	\na	$
\\ 
$Z \gamma \rightarrow  H H $ &
	$	\na	$&$	\na	$&$	\na	$&$	\na	$&$	\na	$&$	-\frac{\cw}{2\sw}	$&    
   	$	\na	$&$	\na	$&$	\na	$&$	\na	$&$	\na	$&$	\frac{\sw}{2\cw}	$   
\\ 
$\gamma \gamma \rightarrow  H H $ &
	$	\na	$&$	\na	$&$	\na	$&$	\na	$&$	\na	$&$	-\frac{1}{2}	$&    
   	$	\na	$&$	\na	$&$	\na	$&$	\na	$&$	\na	$&$	-\frac{1}{2}	$   
\\ 
$Z H \rightarrow  Z H $ &
	$	\na	$&$	\na	$&$	\na	$&$	\na	$&$	-\frac{\cwsq}{4\swsq}X	$&$	\na	$&
   	$	\na	$&$	\na	$&$	\na	$&$	\na	$&$	-\frac{\swsq}{4\cwsq}X	$&$	\na	$   
\\  
$\gamma H \rightarrow  \gamma H $ &
	$	\na	$&$	\na	$&$	\na	$&$	\na	$&$	-\frac{1}{4}X	$&$	\na	$&
   	$	\na	$&$	\na	$&$	\na	$&$	\na	$&$	-\frac{1}{4}X	$&$	\na	$   
\\ 
$Z H \rightarrow \gamma H$ &
	$	\na	$&$	\na	$&$	\na	$&$	\na	$&$	-\frac{\cw}{4\sw}X	$&$	\na	$&
   	$	\na	$&$	\na	$&$	\na	$&$	\na	$&$	\frac{\sw}{4\cw}X	$&$	\na	$  \\
\hline
\end{tabular}
\caption{Same as Table~\ref{tab:fw} for the operators ${\cal O}_{WW}$
  and ${\cal O}_{BB}$.\label{tab:fwwfbb}}
\end{table}
\begin{table}[h]
\begin{tabular}{|c||c|c|c|c|c|c|c|c|}
\hline
 & \multicolumn{8}{c|}{$(\times 2 e^4 \frac{f_{WWW}}{\Lambda^2}\times s)$}  
\\ \hline & $00++$ & $0+0-$ & $0+-0$ & $+00-$ & $+0-0$ & $++00$ & 
$\begin{array}{c}
+++-\\
+---\\
+-++\\
++-+\end{array}$
&$++- -$\\ 
\hline
$W^+W^+ \rightarrow W^+W^+$&$	0	$&$	-\frac{3(2+Y)}{32\swf}	$&$	\frac{3(2+X)}{32\swf}	$&$	\frac{3(2+X)}{32\swf}	$&$	
-\frac{3(2+Y)}{32\swf}	$&$	0	$&
	$	-\frac{3}{4\swf}	$&$\frac{3}{2\swf}$	\\ 
\hline
$W^+Z \rightarrow W^+Z$&$	\frac{3(Y-X)\cw}{32\swf}	$&$	0	$&$	\frac{3(X+2)\cw}{32\swf}	$&$	\frac{3(X+2)\cw}{32\swf}	$&$	0	$&$	\frac{3(Y-X)\cw}{32\swf}	$&
	$	-\frac{3\cwsq}{8\swf}X	$&$	\frac{3\cwsq}{4\swf}X	$
\\ 
$W^+\gamma \rightarrow W^+\gamma$&$	\na	$&$	0	$&$	\na	$&$	\na	$&$	\na	$&$	\na	$&
	$	-\frac{3}{8\swsq}X	$&$	\frac{3}{4\swsq}X	$\\
$W^+Z \rightarrow W^+\gamma$&$	-\frac{3(Y-X)}{32\swt}	$&$	0	$&$	\na	$&$	\frac{3(X+2)}{32\swt}$&$	\na	$&$	\na	$&
	$	-\frac{3\cw}{8\swt}X	$&$	\frac{3\cw}{4\swt}X	$\\
$W^+ Z \rightarrow W^+ H$&$	\na	$&$	\na	$
&$	\frac{3(X+2)\cw}{32\swf}	$&$	\na	$
&$	\frac{3(2+Y)}{32\swf}	$&$	-\frac{3(Y-X)\cw}{32\swf}	$&
	$	\na	$&$	\na	$\\
$W^+ \gamma \rightarrow W^+ H$ &$	\na	$&$	\na	$
&$	\frac{3(X+2)}{32\swt}	$&$	\na	$&$	\na	$
&$	-\frac{3(Y-X)}{32\swt}	$&
	$	\na	$&$	\na	$\\
 \hline
$W^+W^- \rightarrow W^+W^-$&	$\frac{3(Y-X)}{32\swf}	$
&$	\frac{3(2+Y)}{32\swf}	$&$	0	$&$	0	$
&$	\frac{3(2+Y)}{32\swf}	$&$	\frac{3(Y-X)}{32\swf}	$&
	$	\frac{3}{8\swf}Y	$&$	-\frac{3}{4\swf}Y	$\\
$W^+W^- \rightarrow Z Z$ &$	0	$&$	\frac{3(2+Y)\cw}{32\swf}	$&$	-\frac{3(X+2)\cw}{32\swf}	$&$	-\frac{3(X+2)\cw}{32\swf}	$&$	\frac{3(2+Y)\cw}{32\swf}	$&$	0	$&
	$	\frac{3\cwsq}{4\swf}	$&$	-\frac{3\cwsq}{2\swf}	$\\
$W^+W^- \rightarrow \gamma\gamma$ &$	0	$&$	\na	$&$	\na	$&$	\na	$&$	\na	$&$	\na	$&
	$	\frac{3}{4\swsq}	$&$	-\frac{3}{2\swsq}	$\\
$W^+W^- \rightarrow Z \gamma$ &$	0	$&$	\frac{3(2+Y)}{32\swt}	$&$	\na	$&$	-\frac{3(2+X)}{32\swt}	$&$	\na	$&$	\na	$&
	$	\frac{3\cw}{4\swt}	$&$	-\frac{3\cw}{2\swt}	$\\
$W^+W^- \rightarrow Z H$ &$	\na	$&$	\na	$
&$	-\frac{3(2+X)\cw}{32\swf}	$&$	\na	$
&$	-\frac{3(2+Y)\cw}{32\swf}	$&$	\frac{3(Y-X)}{32\swf}	$&
	$	\na	$&$	\na	$\\
$W^+W^- \rightarrow \gamma H$ &$	\na	$&$	\na	$
&$	-\frac{3(X+2)}{32\swt}	$&$	\na	$
&$	-\frac{3(2+Y)}{32\swt}	$&$	\na	$&
	$	\na	$&$	\na	$\\
\hline
\end{tabular}
\caption{Same as Table~\ref{tab:fw} for the operator ${\cal O}_{WWW}$.  
\label{tab:fwww}}
\end{table}
\begin{table}[h]
\begin{tabular}{|l|c|l|}
\hline 
Process  & $\sigma_1,\sigma_2,\lambda_3,\lambda_4$   &  Amplitude \\ \hline
$e^+e^-\to W^-W^+$ 	&	$-+00$	&$	-\frac{i g^2 s \sin \theta}{8}\frac{\cwsq\fw+\swsq\fb}{\cwsq\Lambda^2}		$\\
					&	$+-00$	&$	-\frac{i g^2 s \sin \theta}{4}\frac{\swsq\fb}{\cwsq\Lambda^2}			$\\
					&	$-+- -$	&$	-\frac{3ig^4 s \sin \theta}{8}\frac{f_{WWW}}{\Lambda^2}$\\
					&	$-+++$	&$	-\frac{3ig^4 s \sin \theta}{8}\frac{f_{WWW}}{\Lambda^2}$\\
					\hline
$\nu\bar{\nu}\to W^-W^+$:&	$-+00$	&$	\frac{i g^2 s \sin \theta}{8}\frac{\cwsq\fw-\swsq\fb}{\cwsq}		$\\
					&	$+-00$	&$	0											$\\
					&	$-+- -$	&$	\frac{3ig^4 s \sin \theta}{8}\frac{f_{WWW}}{\Lambda^2}$\\
					&	$-++ +$	&$	\frac{3ig^4 s \sin \theta}{8}\frac{f_{WWW}}{\Lambda^2}$\\
					\hline
$u\bar{u}\to W^-W^+$	&	$-+00$	&$	\frac{i g^2 \nc s \sin \theta}{8}\frac{3\cwsq\fw+\swsq\fb}{3\cwsq}	$\\
					&	$+-00$	&$	\frac{i g^2 \nc s \sin \theta}{6}\frac{\swsq}{\cwsq}\fb			$\\
					&	$-+- -$	&$	\frac{3ig^4\nc s \sin \theta}{8}\frac{f_{WWW}}{\Lambda^2}$\\
					&	$-+++$	&$	\frac{3ig^4\nc s \sin \theta}{8}\frac{f_{WWW}}{\Lambda^2}$\\
					\hline
$d\bar{d}\to W^-W^+$	&	$-+00$	&$	-\frac{i g^2 \nc s \sin \theta}{8}\frac{3\cwsq\fw-\swsq\fb}{3\cwsq}	$\\
					&	$+-00$	&$	-\frac{i g^2 \nc s \sin \theta}{12}\frac{\swsq\fb}{\cwsq\Lambda^2}		$\\
					&	$-+- -$	&$	-\frac{3ig^4\nc s \sin \theta}{8}\frac{f_{WWW}}{\Lambda^2}$\\
					&	$-+++$	&$	-\frac{3ig^4\nc s \sin \theta}{8}\frac{f_{WWW}}{\Lambda^2}$\\
					\hline
$e^+\bar{\nu}\to W^+Z$	&	$-+00$	&$	\frac{i g^2 s \sin \theta}{4\sqrt{2}}\frac{\fw}{\Lambda^2}$\\
					&	$+-00$	&$	0$\\
					&	$-+- -$	&$	\frac{3i\cw g^4 s \sin \theta}{4\sqrt{2}}\frac{f_{WWW}}{\Lambda^2}$\\
					&	$-+++$	&$	\frac{3i\cw g^4 s \sin \theta}{4\sqrt{2}}\frac{f_{WWW}}{\Lambda^2}$\\
					\hline
$e^+\bar{\nu}\to W^+A$:	&	$-+00$	&$	0$\\
					&	$+-00$	&$	0$\\
					&	$-+- -$	&$	\frac{3i\sw g^4 s \sin \theta}{4\sqrt{2}}\frac{f_{WWW}}{\Lambda^2}$\\
					&	$-+++$	&$	\frac{3i\sw g^4 s \sin \theta}{4\sqrt{2}}\frac{f_{WWW}}{\Lambda^2}$\\
\hline 					
\end{tabular}
\caption{ Unitarity violating (growing as $s$) terms of the scattering
  amplitudes $\mathcal{M}({f_1}_{\sigma_1}{\bar{f_2} }_{\lambda_2} \to
  {V_3}_{\lambda_3}{V_4}_{\lambda_4})$ for fermions and gauge bosons
  with the helicities $\sigma_1\sigma_2\lambda_3\lambda_4$ given in
  the second column.
\label{fermiontable}}
\end{table}


\begin{thebibliography}{9}


\bibitem{Bilchak:1987cp} 
  C.~Bilchak, M.~Kuroda and D.~Schildknecht,
  Nucl.\ Phys.\ B {\bf 299}, 7 (1988).
%
\bibitem{Gounaris:1993fh} 
  G.~J.~Gounaris, J.~Layssac and F.~M.~Renard,
  Phys.\ Lett.\ B {\bf 332}, 146 (1994)
  [hep-ph/9311370].
%
\bibitem{Gounaris:1994cm} 
  G.~J.~Gounaris, J.~Layssac, J.~E.~Paschalis and F.~M.~Renard,
  Z.\ Phys.\ C {\bf 66}, 619 (1995)
  [hep-ph/9409260].
%
\bibitem{Gounaris:1995ed}
  G.~J.~Gounaris, F.~M.~Renard and G.~Tsirigoti,
  Phys.\ Lett.\ B {\bf 350} (1995) 212
  [hep-ph/9502376].
%
\bibitem{Degrande:2013mh} 
  C.~Degrande,
  EPJ Web Conf.\  {\bf 49}, 14009 (2013)
  [arXiv:1302.1112 [hep-ph]].
%
\bibitem{Baur:1987mt} 
  U.~Baur and D.~Zeppenfeld,
  Phys.\ Lett.\ B {\bf 201}, 383 (1988).
\bibitem{59ops}
  W.~Buchmuller and D.~Wyler,
  Nucl.\ Phys.\ B {\bf 268}, 621 (1986);
  B.~Grzadkowski, M.~Iskrzynski, M.~Misiak and J.~Rosiek,
  JHEP {\bf 1010}, 085 (2010)
  [arXiv:1008.4884 [hep-ph]].
%
\bibitem{HISZ} 
 K.~Hagiwara, S.~Ishihara, R.~Szalapski and D.~Zeppenfeld,
  Phys.\ Rev.\ D {\bf 48}, 2182 (1993).
%
\bibitem{HISZ2} 
 K.~Hagiwara, T.~Hatsukano, S.~Ishihara and R.~Szalapski,
  Nucl.\ Phys.\ B {\bf 496}, 66 (1997)
  [hep-ph/9612268].
%
\bibitem{us1} 
  T.~Corbett, O.~J.~P.~Eboli, J.~Gonzalez-Fraile and M.~C.~Gonzalez-Garcia,
  Phys.\ Rev.\ D {\bf 86}, 075013 (2012)
  [arXiv:1207.1344 [hep-ph]].
%
\bibitem{us2} 
  T.~Corbett, O.~J.~P.~Eboli, J.~Gonzalez-Fraile and
  M.~C.~Gonzalez-Garcia, 
  Phys.\ Rev.\ D {\bf 87}, 015022
  (2013) [arXiv:1211.4580 [hep-ph]]. 
%
\bibitem{us3}
  T.~Corbett, O.~J.~P.~Eboli, J.~Gonzalez-Fraile and M.~C.~Gonzalez-Garcia,
  Phys.\ Rev.\ Lett.\  {\bf 111}, no. 1, 011801 (2013)
  [arXiv:1304.1151 [hep-ph]].
%
\bibitem{Brivio:2014pfa} 
  I.~Brivio, O.~J.~P.~Eboli, M.~B.~Gavela, M.~C.~Gonzalez-Garcia, L.~Merlo and S.~Rigolin,
  arXiv:1405.5412 [hep-ph].
%
\bibitem{Hagiwara:1986vm}
 K.~Hagiwara, R.~D.~Peccei, D.~Zeppenfeld and K.~Hikasa,
  Nucl.\ Phys.\ B {\bf 282}, 253 (1987).
%
\bibitem{Hagiwara:1993ck}
K.~Hagiwara, S.~Ishihara, R.~Szalapski, and D.~Zeppenfeld,
Phys.\ Rev.\ {\bf D48}, 2182 (1993).
%
\bibitem{Alam:1997nk}
  S.~Alam, S.~Dawson and R.~Szalapski,
  Phys.\ Rev.\ D {\bf 57}, 1577 (1998)
  [hep-ph/9706542].
%
\bibitem{DeRujula:1991se}
A.~De~Rujula, M.~Gavela, P.~Hernandez, and E.~Masso,
Nucl.\ Phys.\ {\bf B384}, 3 (1992).
%
\bibitem{Baak:2014ora} 
  M.~Baak {\it et al.}  [Gfitter Group Collaboration],
  Eur.\ Phys.\ J.\ C {\bf 74}, 3046 (2014)
  [arXiv:1407.3792 [hep-ph]].
%
\bibitem{Jacob:1959at} 
  M.~Jacob and G.~C.~Wick,
  Annals Phys.\  {\bf 7}, 404 (1959)
  [Annals Phys.\  {\bf 281}, 774 (2000)].
%
\bibitem{Csaki:2003dt} 
  C.~Csaki, C.~Grojean, H.~Murayama, L.~Pilo and J.~Terning,
  Phys.\ Rev.\ D {\bf 69}, 055006 (2004)
  [hep-ph/0305237].
%
\bibitem{pdg} 
  K.~A.~Olive {\it et al.}  [Particle Data Group Collaboration],
  Chin.\ Phys.\ C {\bf 38}, 090001 (2014).


\end{thebibliography}
\end{document}